\documentclass[%
 reprint,
superscriptaddress,
showpacs,
 amsmath,amssymb,
 aps,
]{revtex4-1}

\usepackage{graphicx}
\usepackage{dcolumn}
\usepackage{bm}

\usepackage{siunitx} 

\begin{document}

\preprint{APS/123-QED}

\title{Bias current dependence of superconducting transition temperature in superconducting spin valve nanowires}

\author{Alejandro A. Jara}%
 \email{jaraabaa@uci.edu} 
\affiliation{Department of Physics and Astronomy, University of California, Irvine, California 92697, USA}
\affiliation{Departamento de F\'isica, Universidad T\'ecnica Federico Santa Mar\'ia, Valpara\'iso 2390123, Chile}

\author{Evan Moen}
\email{moenx359@umn.edu} 
\author{Oriol T. Valls}
\email{otv@umn.edu} 
\affiliation{School of Physics and Astronomy, University of Minnesota, Minneapolis, Minnesota 55455, USA}

\author{Ilya N. Krivorotov}
\email{ilya.krivorotov@uci.edu}
\affiliation{Department of Physics and Astronomy, University of California, Irvine, California 92697, USA}

\date{November 25, 2019}

\begin{abstract}
Competition between superconducting and ferromagnetic ordering at interfaces between ferromagnets (F) and superconductors (S) gives rise to several proximity effects such as odd-triplet superconductivity and spin-polarized supercurrents. A prominent example of an S/F proximity effect is the spin switch effect (SSE) observed in S/F/N/F superconducting spin valve multilayers, in which the superconducting transition temperature T$_c$ is controlled by the angle $\phi$ between the magnetic moments of the F layers separated by a nonmagnetic metallic spacer N. Here we present an experimental study of SSE in Nb/Co/Cu/Co/CoO$_x$ nanowires measured as a function of bias current flowing in the plane of the layers. These measurements reveal an unexpected dependence of T$_c(\phi)$ on the bias current:  T$_c(\pi)$--T$_c(0)$ changes sign with increasing current bias. We attribute the origin of this bias dependence of the SSE to a spin Hall current flowing perpendicular to the plane of the multilayer, which suppresses T$_c$ of the multilayer. The bias dependence of SSE can be important for hybrid F/S devices such as those used in cryogenic memory for superconducting computers as device dimensions are scaled down to the nanometer length scale.

\end{abstract}

\maketitle

\section{\label{sec:level1}Introduction}

Superconducting computing is an active area of research. At present, two major directions of superconducting computing are actively pursued. First, quantum computers based on Josephson junction (JJ) qubits are being built and tested by multiple research groups \cite{Nakamura1999, Bergeal2010, Paik2011, Lucero2012, Barends2013, Devoret2013, Krantz2019, Semenov2019}. Second, classical cryogenic computers based on single flux quantum (SFQ) logic offer significant advantages in speed and energy efficiency over their classical semiconductor-based counterparts \cite{Holmes2013, Manheimer2015, Holmes2015, Larkin2012}. One roadblock for the SFQ computing is the absence of a scalable energy-efficient memory that is impedance matched to the low-resistance SFQ JJ-based logic \cite{Holmes2013}. In rapid SFQ (RSFQ) devices, static power consumed by memory can exceed the dynamic power required for logic operation by two orders of magnitude. One potential solution to this problem is the use of all-metallic F/N/F spin valves (SVs) that consist of two F layers separated by a non-magnetic metallic spacer N. To form a non-volatile magnetic memory element, such a SV can be incorporated as a magnetic barrier in a JJ \cite{Bergeret2001, Blanter2004, Linder2007, Baek2015,Schneider2018, Dayton2018}. Switching of the relative orientation of the F layer's magnetic moments modifies the JJ critical current. For this type of memory, the SV is in direct electrical contact with a superconducting film, and thus understanding of thermodynamic and magneto-transport properties of S/F/N/F heterostructures is important for the design of such memory elements.

An F/N/F spin valve exhibits the giant magneto-resistance (GMR) effect, in which the resistance of the multilayer depends on the angle $\phi$ between magnetic moments of the two F layers \cite{Dieny1991}. The parallel (P) configuration of the magnetic moments usually has lower resistance than the antiparallel (AP) configuration, R$_P$ $<$ R$_{AP}$. When a SV is interfaced with an S layer, the magneto-resistance (MR) of the S/F/N/F multilayer can strongly differ from GMR in F/N/F for temperatures close to the superconducting transition temperature T$_c$. The sign of the MR near T$_c$ can be opposite to that of GMR well above T$_c$ \cite{Gu2002, Potenza2005, Moraru2006, Miao2008, Zhu2009, zkhv, alejandro}. This dependence of T$_c$ on the magnetic configuration, known as the spin switch effect (SSE), is a result of magnetic control of the superconducting proximity effect where the degree of the condensate penetration into the magnetic layer is determined by the relative weight of the singlet and exchange-field-induced odd-triplet contributions to the Cooper pair wave function, which is dependent on $\phi$. In general, $\Delta$T$_c$ $\equiv$ T$_c$(AP)--T$_c$(P) is an oscillatory function of the F layer thickness due to quantum interference effects in the magnetic multilayer \cite{Rusanov2006, Steiner2006, Stamopoulos2007, Singh2007, Zhu2009, alejandro}.
 
 \begin{figure*}[pt]
\includegraphics[width=0.99\textwidth]{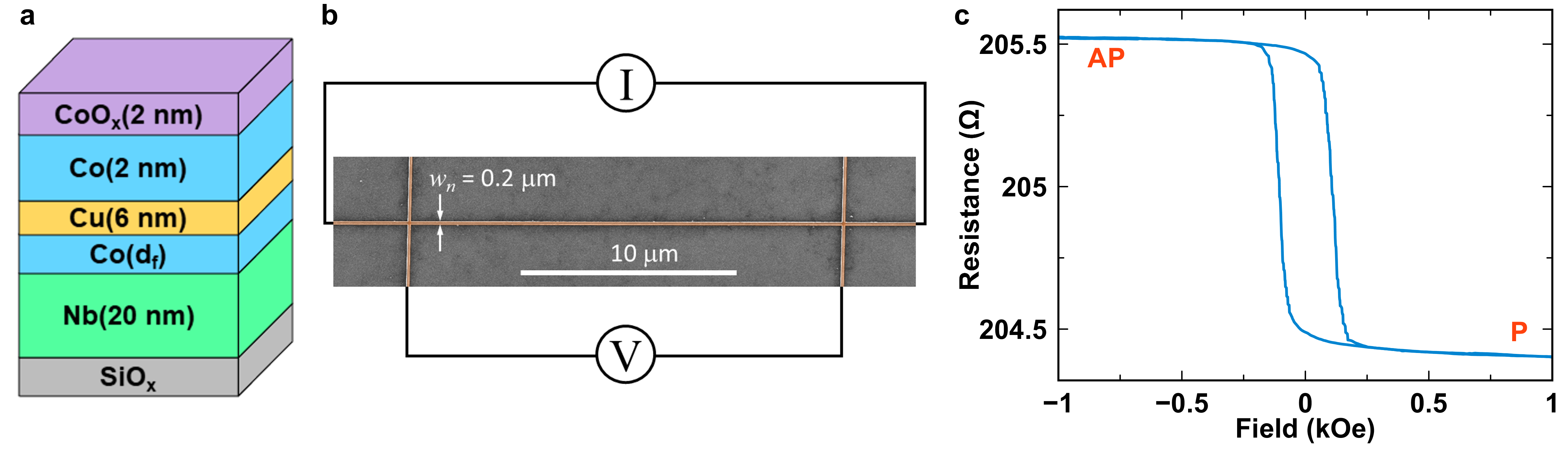}
\caption{(\textbf{a}) Schematic of the superconducting spin valve multilayer. (\textbf{b}) Scanning electron micrograph of the nanoscale superconducting spin valve Hall bar. Nanowire length between the Hall crosses L = 20 $\mu$m, nanowire width w$_n$ = 200 nm. (\textbf{c}) Magnetoresistance of the Nb(20 nm)/Co(0.7 nm)/Cu(6 nm)/Co(2 nm)/CoO$_x$(2 nm) device shown in (\textbf{b}) measured at T = 10 K for magnetic field applied parallel to the nanowire axis.}
\label{Fig:Schematic_RvsH}
\end{figure*}
 
The SSE has been extensively studied over the past decade \cite{Zhu2009, alejandro, Leksin2010, Leksin2011, Leksin2012, Wang2014}. Previous experimental studies concentrated on measurements of SSE in thermodynamic equilibrium where small electrical currents are used to probe the T$_c$ and MR of the system \cite{alejandro, Leksin2010, Wang2014}. However, when S/F/N/F structures are used as core components of nanoscale non-volatile memory, they are exposed to relatively high current densities which are needed to achieve high signal-to-noise ratio in a superconducting circuit. Such high current densities can affect both T$_c$ and MR of these structures. In this paper, we present an experimental study of the SSE in Nb/Co/Cu/Co/CoO$_x$ nanowires as a function of current bias. We find an unexpectedly strong dependence of $\Delta$T$_c$ on bias current where $\Delta$T$_c$ reverses sign with increasing current. We attribute (see below) the origin of this effect to a spin Hall current flowing perpendicular to the plane of the multilayer, which suppresses the critical temperature. We rule out other possible explanations on either qualitative or quantitative grounds.

\section{\label{sec:level2} Experimental Results} 

Fabrication of the S/F/N/F nanowire devices begins by deposition of a series of (substrate)/Nb(20 nm)/Co(d$_f$)/Cu(6 nm)/Co(2 nm)/CoO$_x$(2 nm) multilayers by magnetron sputtering onto thermally oxidized Si wafers, as illustrated in Fig.~\ref{Fig:Schematic_RvsH}a. The thickness $d_f$ of the bottom Co layer varies in the range of $0.6-0.7$ nm. The multilayers are deposited at room temperature in 2 mTorr of Ar process gas in a high vacuum system with a base pressure below $10^{-8}$ Torr. The 2 nm thick CoO$_x$ layer is formed via natural oxidation of the top Co layer in air \cite{Tompkins1981}. The multilayers are patterned into nanoscale Hall bars via electron-beam lithography using ma-N 2401 negative e-beam resist and subsequent Ar ion mill etching. Figure \ref{Fig:Schematic_RvsH}b shows a scanning electron micrograph of the device and its dimensions: the central part of the Hall bar is a 200 nm wide, 20 $\mu$m long Nb(20 nm)/Co(d$_f$)/Cu(6 nm)/Co(2 nm)/CoO$_x$(2 nm) multilayer nanowire.

\begin{figure*}[pt]
\includegraphics[width=0.99\textwidth]{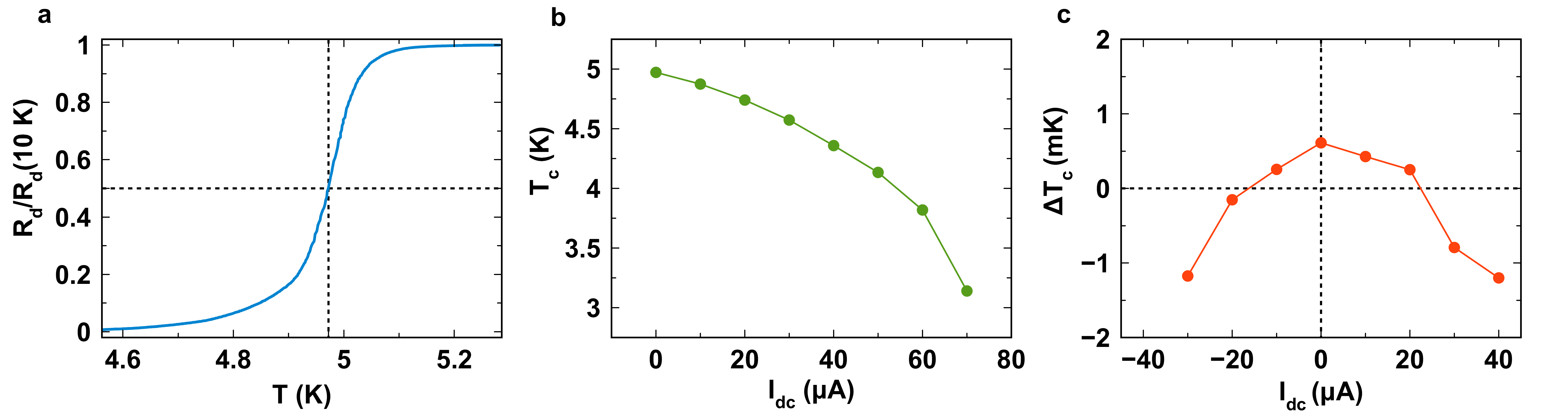}
\caption{(\textbf{a}) Differential resistance R$_d$ of the Nb(20 nm)/Co(0.7 nm)/Cu(6 nm)/Co(2 nm)/CoO$_x$(2 nm) nanowire device measured as a function of temperature and normalized to its value at 10 K. Dependence of (\textbf{b}) the T$_c$ in the P state and (\textbf{c}) $\Delta$T$_c$ = T$_c$(AP)--T$_c$(P) on direct current bias I$_{dc}$ applied to the nanowire.}
\label{Fig:DTcvsI_RdvsT}
\end{figure*}

At cryogenic temperatures, the direction of magnetization of the top Co layer is pinned by a strong exchange bias field from the antiferromagnetic CoO$_x$ layer \cite{Gredig2000,Gredig2002}. In contrast, the magnetization of the bottom Co layer is easily saturated by a small magnetic field, as we show below. For all our measurements, the exchange bias field direction is set by a 3 kOe in-plane magnetic field applied along the nanowire axis when the sample is cooled from room temperature to 10 K. We characterize electrical transport in the nanowires by using four-point resistance (R) and four-point differential resistance (R$_d$ = dV/dI) measurements. For resistance measurements, we employ a direct current source and a digital nanovoltmeter.  Differential resistance measurements are made by the lock-in technique using a 1 $\mu$A alternating current source. All-electric transport measurements are made in a continuous flow $^4$He cryostat with temperature stability of 0.1 mK at temperatures below 15 K. 
 
At temperatures above T$_c$, all samples exhibit conventional current-in-plane GMR. Fig.~\ref{Fig:Schematic_RvsH}c shows the resistance of the Nb(20 nm)/Co(0.7 nm)/Cu(6 nm)/Co(2 nm)/CoO$_x$(2 nm) nanowire measured as a function of magnetic field applied parallel to the nanowire axis, at T = 10 K $>$ T$_c$. This GMR curve reveals that the magnetization the free Co layer can be switched between parallel (P) and antiparallel (AP) orientations with respect to the magnetization of the pinned Co layer by a small magnetic field of 0.2 kOe. Since the GMR hysteresis loop in Fig.~\ref{Fig:Schematic_RvsH}c is not shifted from zero along the magnetic field axis, interlayer exchange coupling between the Co layers across the nonmagnetic Cu spacer is negligibly small \cite{alejandro}. The magnitude of the GMR is relatively small (0.55$\,\%$) due to significant electric current shunting through the Nb layer.

Fig.~\ref{Fig:DTcvsI_RdvsT}a shows the differential resistance of the nanowire versus temperature measured at zero direct current bias. This figure reveals that T$_c\approx 5$\,K is significantly reduced compared to that of bulk Nb due to the pair-breaking exchange field from the proximate Co layer \cite{Eschrig2011}. In this paper, T$_c$ is defined as the midpoint of the resistor-to-superconductor transition:  R$_d$(T$_c$)=R$_d$(10\,K)/2. Application of direct current to the nanowire reduces T$_c$ due to the orbital pair breaking effect as illustrated in Fig.~\ref{Fig:DTcvsI_RdvsT}b. 

On these nanowire samples, we then measure the current bias dependence of the spin switch effect. In these measurements, we fix the direct current I$_{dc}$ flowing through the nanowire and slowly sweep the sample temperature through T$_c$ at the rate of 2 mK per minute. Throughout this temperature sweep, we alternate external the magnetic field along the wire between +1 kOe and --1 kOe in order to switch the sample between the P and AP states and measure R$_d$ in these states. Using the R$_d$(T) curves given by these measurements for the P and AP configurations, we can extract  $\Delta$T$_c \equiv$ T$_c$(AP)-T$_c$(P) at a given value of the current bias. The measured dependence of $\Delta$T$_c$ on the current bias for the  Nb(20 nm)/Co(0.7 nm)/Cu(6 nm)/Co(2 nm)/CoO$_x$(2 nm) nanowire is shown in Fig.~\ref{Fig:DTcvsI_RdvsT}c. This figure reveals that $\Delta$T$_c$ decreases with increasing $|$I$_{dc}|$ and changes sign near $|$I$_{dc}|=20$\,$\mu$A. Measurements of $\Delta$T$_c$ versus I$_{dc}$ made for samples with different values of the free layer thickness d$_f$ reveal a similar trend as shown in Fig.~\ref{Fig:DTcvsI_All}. It is interesting to note that $|\Delta$T$_c|$ at a non-zero current bias can be significantly enhanced compared to its zero-bias value. This current-induced enhancement of the spin switch effect magnitude may find use in applications.

\section {Discussion and Interpretation} 
\label{theory}

An understanding of the results given above can be gained by considerations based on previous theoretical  work. The thermodynamic properties of S/F layered systems considered here have been quantitatively studied using the methods in Refs.~\cite{Halterman2002,hbv,hoprl}. The self consistent methodology developed there has been used to explain the equilibrium behavior of similarly fabricated samples~\cite{zkhv} and, significantly, to explain in quantitative detail the behavior of T$_c$ in F$_1$/N/F$_2$/S spin valve samples~\cite{alejandro}, such as those described in the previous section, as a function of misalignment angle $\phi$  ($\phi =0$ for P configuration and $\phi=\pi$ for AP configuration), and of the thickness of the different layers. This agreement was achieved with material parameters appropriate to Co for the F layers and Nb for the superconductor. The agreement in detail between theory and experiment in that work gives us reasonable confidence in the ability of the theory to give an explanation of the observed experimental results. 

\begin{figure}[pt]
\includegraphics[width=0.99\columnwidth]{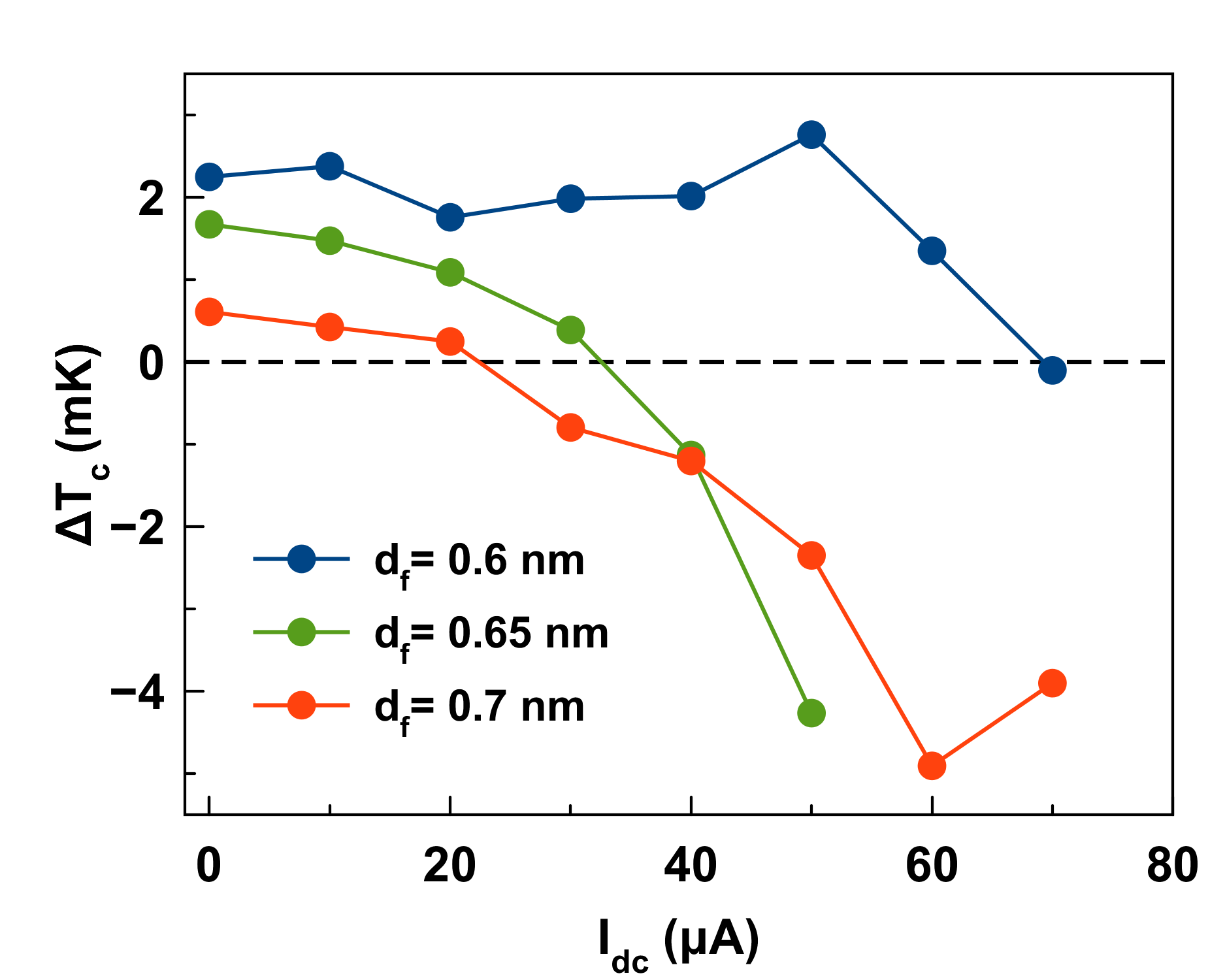}
\caption{Dependence of $\Delta$T$_c \equiv$ T$_c$(AP)--T$_c$(P) on current bias I$_{dc}$ for three Nb(20 nm)/Co(d$_f$)/Cu(6 nm)/Co(2 nm)/CoO$_x$(2 nm) devices with different values of the free layer thickness d$_f$.}
\label{Fig:DTcvsI_All}
\end{figure}

These considerations apply also to the samples studied here. They allow us to fulfill the objective of proposing a plausible quantitative understanding of the data which, while unusual, is reliable and reproducible. To see how to proceed we consider the experimental data for the equilibrium quantity $\Delta$T$_c$ for our samples, as plotted in Fig.~\ref{Fig:DTcvsI_All}. This
quantity decreases with $d_f$ in the plot. In general it is an oscillatory~\cite{alejandro} function of $d_f$, having a period determined \cite{Moen2018split,demler} by the internal field of the magnet, in this case cobalt. We have verified that in the range of $d_f$ plotted in this figure, the theory predicts that $\Delta$T$_c$ decreases with $d_f$, that is, the $d_f$ thicknesses are in the decreasing portion of one of the oscillations. The theoretically calculated $\Delta T_c$ for the respective $d_f$ values of these samples quantitatively agree with the experimental results, as was the case for the samples in Ref.~\onlinecite{alejandro}. From this it follows that we can be highly confident that we can profitably use insights from
the theory to interpret the transverse current results. This will be done in the rest of this Section.

We turn next to the dependence on I$_{dc}$, in particular, the dependence of $\Delta$T$_c$ on the transverse (flowing in the plane of the layers) current as depicted in  Fig.~\ref{Fig:DTcvsI_RdvsT}c. At finite current, this is not an equilibrium quantity. Previous theoretical work on these devices, which we will use here, has been performed~\cite{wvhg,Moen2017,Moen2018,Moen2018split} only for the case where the current is longitudinal, i.e. perpendicular to the sample's layers. Although this work yields, as we shall see, very useful insights into the situation discussed here, our discussion will necessarily be semi-quantitative only for the current in-plane measurements. However, our analysis does exclude, we believe conclusively, many possibilities, and points to a likely explanation in terms of the spin Hall effect \cite{Kontani2009,Takahashi2012,Espedal2017,Derunova2019, Wakamura2015}.

The first guess that one might make when examining the data is that the change in the relative T$_c$ is due to curves of T$_c$ vs I$_{dc}$ reflecting different correlations between these quantities in the P and AP states. At higher current density, the critical temperature decreases, as shown in Fig.~\ref{Fig:DTcvsI_RdvsT}b. The difference between the P and AP critical temperatures may increase as the current density approaches the critical value  where T$_c$ goes to zero. However, what is not expected is for the relative critical temperature to change sign between P and AP states: one state should always have a greater T$_c$ than the other as they approach the critical current density. We rule out this simple explanation for the $\Delta$T$_c$ features described.

We turn now to explanations in terms of more exotic transport phenomena. We first consider the ordinary Seebeck effect. This would require the existence of a temperature gradient across the S layer. It is certainly possible (and we will consider this possibility in connection with the {\it spin} Seebeck effect below) that there is a temperature gradient of unknown magnitude between the top and bottom of the nanowire. It does not seem  reasonable, however, that there should be a temperature gradient across the entire nanowire in-plane with the layers. From these considerations, we can rule out the ordinary Seebeck effect, as there is no temperature gradient along the direction of the applied electric field that would affect the measurement. Furthermore, the magnetizations of the ferromagnetic layers are always collinear with the current applied for the P and AP states. This rules out also the Hall effect (strictly speaking the anomalous Hall effect) and also the Nernst effect. This leaves us with the spin Hall and spin Seebeck effects as being geometrically possible phenomena to explain the experimental data.

Consider first the spin Hall effect, that is, the creation of a spin current normal to a charge current. Significant spin Hall current density can be generated by Nb at T$_c$. Indeed, thermal fluctuations at T$_c$ render the Nb layer a fluctuating mix of normal and superconducting regions. The usual intrinsic and extrinsic mechanisms of spin Hall current generation are active within the normal regions. Also, spin Hall current generated in the normal regions can be efficiently transported through the superconducting regions at T$_c$. Recent theoretical work~\cite{Taira2018} demonstrates that spin diffusion length in superconductors near T$_c$ is greatly enhanced compared to that in the normal state. Therefore, the Nb layer can both generate and transport spin Hall current at T$_c$. In our argument, we are looking for a possible relationship between the observed transition temperature difference $\Delta$T$_c$ and the current applied I$_{dc}$ in the transverse direction within the superconductor. From our transport calculations~\cite{Moen2017,wvhg}, we have found that the relative  conductance, calculated for a longitudinal current between P and AP states can shift abruptly near the critical bias. We also have demonstrated~\cite{Moen2018split} that this phenomenon  arises from the marked difference between conductances in the up and down spin channels, which interact differently with the aligned or misaligned magnetizations. We then consider whether the longitudinal spin current, which we call I$_S$, created by  the transverse current I$_{dc}$ via the spin Hall effect can produce an effect on $\Delta$T$_c$ of the relevant order of magnitude. This energy scale of the relative transition temperature should be similar to the energy scale of the critical bias around which we generally see an abrupt shift in the relative charge and spin current between the P and AP states from our previous transport studies~\cite{Moen2018, Moen2018split}.

The Spin Hall effect is characterized by the relation~\cite{shall} 
\begin{equation}
\mathbf{J}_S=\frac{\hbar}{2e}\Theta_{SH}\mathbf{J}_c\times \mathbf{\sigma}
\end{equation} 
where $J_c$ is the charge current density, $J_S$ is the spin current density and $\Theta_{SH}$ a dimensionless coefficient that characterizes the strength of the effect. We take  $\Theta_{SH}\approx10^{-3}$ as given in Ref.~\cite{shall} for Nb. The cross-sectional dimensions of the nanowire are (see Fig.~\ref{Fig:Schematic_RvsH}b) 200 nm by 20 $\mu$m in the direction corresponding to a longitudinal current, while the cross-section of the nanowire is 200 nm by $\sim$30 nm for the transverse current (the effective cross-section may be smaller well below T$_c$, but this does not affect our order-of-magnitude estimates near the transition). In our case, the assumed spin current is in the longitudinal direction while the charge current is in the transverse direction within the nanowire. We find, using these considerations and values, that the spin current I$_S$ $\approx$ I$_{dc}$, due to the value of the coefficient compensating for the difference in the cross-sections. Here the currents are given in the natural units of electron spin per unit time and electron charge per unit time, respectively. For the spin Hall effect to be a plausible explanation of the observation, the relevant energy scale caused by this spin current must be on the order of magnitude of the difference in critical temperatures, $k_B \Delta {\rm T}_c\approx 2\times10^{-4}$ meV. Then, I$_S = k_B \Delta {\rm T}_c G N_{ch}$ where $G$ is the conductance per channel, which is of the order of the quantum of conductance,  $G_0\approx4\times10^{-5}$ A/V, and $N_{ch}$ is the number of channels. If we take I$_{dc}=40 \mu$A, which is approximately when the transition occurs in the P and AP critical temperatures, consistent with Fig.~\ref{Fig:DTcvsI_All}, we find that  I$_S\approx 4\times10^{-5}$. To see if this number is plausible, we can then estimate the number of channels through which the spin current moves through, to find that $N_{ch}\approx5\times10^{6}$. We see this value is reasonable if the spin current moves predominantly near the edges of the nanowire. Alternatively, we can get an upper-bound estimate for $N_{ch}$ if we take $\Theta_{SH}\approx10^{-2}$ which would agree with other spin Hall measurements on Nb~\cite{morota}. Furthermore, if we consider stronger interfacial scattering at the interfaces, our estimate on the conductance $G$ decreases by up to a factor of ten. Taking both adjustments into account, the estimated number of channels becomes $N_{ch}\approx5\times10^{8}$. This is closer to the expected number of channels for a 20 $\mu$m by 200 nm area, with the current flowing throughout the entire sample. This would also mean that the longitudinal resistance within the valve is on the order 50 $\mu \Omega$. Alternatively, one can express the results in terms of the resulting spin current density by dividing the obtained value of I$_S$ by the area in the direction corresponding to the longitudinal current. 
 
Similar considerations can be attempted for the spin Seebeck effect (SSE). The spin Seebeck effect is the production of a spin current induced by a temperature gradient along the same direction. One would have to posit a longitudinal temperature gradient. The spin Seebeck effect can be quantified via~\cite{uchida} $\mu_\uparrow - \mu_\downarrow = e S_S \Delta T$ where $\mu_\uparrow - \mu_\downarrow$ is the spin voltage, $\Delta T$ the temperature difference between the top and the bottom of the wire, and $S_S$ is the spin Seebeck coefficient, which is typically\cite{uchida} on the order of 10 ${\mu}$V/K. To estimate the required temperature gradient to induce the observed  effect on the system, we use the assumptions we made for the spin Hall effect, that I$_S\approx$ I$_c$ in the natural units described above. The relevant bias is on the order of the pair potential (i.e., the critical bias $V_c\approx1$ meV), as we see the most significant effects on the relative P and AP features near this bias in all transport measurements~\cite{Moen2018,Moen2018split}. Thus we can say $\mu_\uparrow - \mu_\downarrow \sim V_c \approx 10^{-3}$ V. Then, if we solve for $\Delta T$ we find that the temperature difference required would be on the order of 100 K, which is absurdly too high to be the situation in this experiment. We, therefore, rule out the spin Seebeck on quantitative grounds. The negative conclusion in this paragraph strengthens the positive conclusion in the previous one: it is perfectly possible, as we see, to find a qualitatively possible explanation that collapses under a more quantitative analysis. 

The possible combined influence of the spin Seebeck effect and the {\it inverse} spin Hall effect \cite{Valenzuela2006,Saitoh2006} could also be considered. In this scenario, a longitudinal spin current, arising from the spin Seebeck effect, would induce an excess transverse charge current via the inverse spin Hall effect. This excess current would reduce the critical temperature of the sample. For $\Delta$T$_c$ to change sign, the relative excess current of the P and AP states must cross-over with increasing applied current. Thus, we look again at energy scales near the critical bias. The temperature gradient would induce a transverse electric field $E_{ISHE}$ via the inverse spin Hall effect~\cite{sola}: $S_S {\Delta T}/{L_z} = - E_{ISHE}$ where $L_z\approx20$ nm is the height of the nanostructure and $\Delta T$ is the temperature difference between the top and bottom of the sample. A cross-over between the P and AP state would occur when $E_{ISHE}=V_c/L_c$ where $L_c\approx20 \mu$m is the distance between contacts and $V_c\approx10^{-3}$ V is the critical bias. Thus $\Delta T =({V_c}/{S_S})({L_z}/{L_c})$. For $S_S\approx10^{-5}$ V/K, we get a difference in temperature on the order of 0.1 K. Although the  temperature gradient is reduced significantly, this is still too high to be plausible for the all-metallic multilayer used in our experiment, and we rule out this possibility. 

One can also consider an alternative explanation in terms of a putative consequence of a proximity effect where Cooper pairs would diffuse over a shorter length scale in the AP configuration (i.e. the ``more insulating'' configuration) than in the P configuration. Using the measured values of the conductivity of the Co/Cu/Co spin valve in the AP and P configurations as well as the the conductivity of the Nb layer, we have quantitatively evaluated the contribution of this mechanism to the bias dependence of SSE in our system. In this calculation, we modeled the the nanowire as two parallel resistors: a Nb layer resistor and a Co/Cu/Co layer resistor. When the Co/Cu/Co spin valve layer switches from the P to the AP configuration, its resistance increases and a higher current density flows through the Nb layer. We then used the data in Fig.~\ref{Fig:DTcvsI_RdvsT}b to evaluate how this current density increase changes $T_c$. This calculation, reveals that the bias-induced dependence of $\Delta T_c$ due to this mechanism would be a factor of fifteen smaller than the observed value. Therefore, this alternative mechanism cannot explain our experimental data.

\section {Conclusions}
\label{conclusions}

We have measured the influence of electric current bias on the spin switch effect in Nb/Co/Cu/Co/CoO$_x$ superconducting spin valve nanowires. We observed that  the dependence of the superconducting transition temperature on the transverse electric current (that is, flowing in the plane of the multilayer) is different in the parallel and antiparallel configurations of the spin valve. As a result, the sign of the spin switch effect $\Delta$T$_c \equiv$ T$_c$(AP)--T$_c$(P) and the associated magneto-resistance reverses with increasing current bias. We analyze the origin of the observed effect, and we attribute it to a pure spin Hall current flowing perpendicular to the plane of the layers. The order of magnitude of the observed dependence of $\Delta$T$_c$ on the electric current bias is consistent with the density of spin Hall current in Nb. We discuss several other possible explanations and we conclude  that some of them (such as the ordinary Seebeck and anomalous Hall effects) are qualitatively impossible while some others (e.g., the spin Seebeck effect) are quantitatively implausible. Our work advances the understanding of the physics of proximity effects in ferromagnet/superconductor multilayers away from thermal equilibrium, and it may find use in non-volatile cryogenic memory technology.\\ 

\section {ACKNOWLEDGMENTS}  
This work was supported by DOE Grant No. DE-SC0014467. The computational aspects were supported also by the Minnesota Supercomputer Institute.

\end{document}